\documentclass[conference]{IEEEtran}
\IEEEoverridecommandlockouts
\usepackage[nolist]{acronym} 
\usepackage{cite}
\usepackage{comment}
\usepackage{amsmath,amssymb,amsfonts}
\usepackage{algorithmic}
\usepackage{graphicx}
\usepackage{comment}
\usepackage{textcomp}
\usepackage{xcolor}
\usepackage{wrapfig}
\usepackage{color}
\usepackage{verbatim}
\usepackage{adjustbox}
\usepackage{hyperref}
\usepackage{tabularx}
\usepackage{tikz}

\usepackage{url}

\usepackage{subfigure}
\usepackage{booktabs}
\usepackage{textgreek}
\usepackage{enumitem}
\usepackage[skins,breakable]{tcolorbox} %
\usepackage{multirow}
\usepackage{cprotect}
\usepackage{circledsteps}
\usepackage[perpage]{footmisc}

\usepackage{listings}

\def\BibTeX{{\rm B\kern-.05em{\sc i\kern-.025em b}\kern-.08em
    T\kern-.1667em\lower.7ex\hbox{E}\kern-.125emX}}

\definecolor{circBG}{RGB}{186,90,84} %
\definecolor{circBGBlue}{RGB}{4, 146, 194} %

\DeclareRobustCommand{\circled}[1]{%
  \tikz[baseline=(char.base)]{
    \node[circle, fill=circBG, inner sep=0pt] (char)
    {\color{white}\strut #1};
  }%
}

\DeclareRobustCommand{\circledBlue}[1]{%
  \tikz[baseline=(char.base)]{
    \node[circle, fill=circBGBlue, inner sep=0pt] (char)
    {\color{white}\strut #1};
  }%
}

\newcommand{\openshmem}[1][]{%
 {Open\-SHMEM\ifthenelse{\equal{#1}{}}{}{~#1}}}

\newcommand{\artifacturl}{\url{https://github.com/spcl/zero-sm-artifact}}

\begin{acronym}
\acro{PVM}{Parallel Virutal Machine}
\acro{PICL}{Portable Instrumented Communication Library}
\acro{SSID}{Synchronization Sequence Identification}
\acro{phit}{physical units}
\acro{BTE}{Byte Transfer Engine}
\acro{FMA}{Fast Memory Access}
\acro{MPI}{\emph{Message Passing Interface}}
\acro{UPC}{\emph{Unified Parallel C}}
\acro{PGAS}{Partitioned Global Address Space}
\acro{SIMD}{\emph{Single Instruction Multiple Data}}
\acro{MIMD}{\emph{Multiple Instruction Multiple Data}}
\acro{HPC}{\emph{High Performance Computing}}
\acro{API}{\emph{Application Programming Interface}}
\acro{RDMA}{\emph{Remote Direct Memory Access}}
\acro{AMO}{\emph{Atomic Memory Operations}}
\acro{CRC}{\emph{Cyclic Redundancy Check}}
\acro{MDH}{\emph{Memory Descriptor Handle}}
\acro{PE}{\emph{Programming Element}}
\acro{ORNL}{Oak Ridge National Laboratory}
\acro{UNM}{University of New Mexico}
\acro{MPI}{Message Passing Interface}
\acro{PGAS}{Partioned Global Address Space}
\acro{HPC}{High Performance Computing}
\acro{UCCS}{Universal Common Communication Substrate}
\acro{PROMPI}{Protocol Reconfiguration and Optimization system for \ac{MPI}}
\acro{LANL}{Los Alamos National Laboratory}
\acro{DOE}{Department of Energy}
\acro{DoD}{Department of Defense}
\acro{OLCF}{\emph{Oak Ridge Leadership Computing Facility}}
\acro{QMC}{\emph{Quantum Monte Carlo}}
\acro{CPU}{\emph{Central Processing Unit}}
\acro{SpT}{QMC Samples per Thread}
\acro{AMR}{\emph{Adaptive Mesh Refinement}}
\acro{WL-LSMS}{\emph{Wang Landau - Locally Self-consistent Multiple Scattering}}
\acro{WL}{\emph{Wang Landau}}
\acro{LSMS}{\emph{Locally Self-consistent Multiple Scattering}}
\acro{rLIZ}{\emph{Local Interaction Zone Radius }}
\acro{LAMMPS}{\emph{Large-scale Atomic/Molecular Massively Parallel Simulator}}
\acro{NRDF}{\emph{Non-equilibrium Radiation Diffusion}}
\acro{CESM}{\emph{Community Earth System Model}}
\acro{Vampir}{\emph{Vampir Tool Suite}}
\acro{SAMRAI}{\emph{Structured Adaptive Mesh Refinement Application Infrastructure}}
\acro{CHARMM}{Chemistry at HARvard Macromolecular Mechanics}
\acro{PPPM}{particle-particle particle-mesh}
\acro{IOFSL}{I/O Forwarding Scalability Layer}
\acro{PE}{Processing Element}
\acro{APP}{Application Time}
\acro{COMP}{Computation Time}
\acro{COL}{Collective}
\acro{GPU}{Graphical Processing Unit}
\acro{SIMD}{\emph{Single Instruction Multiple Data}}
\acro{MIMD}{\emph{Multiple Instruction Multiple Data}}
\acro{API}{\emph{Application Programming Interface}}
\acro{RDMA}{\emph{Remote Direct Memory Access}}
\acro{UPC}{\emph{Unified Parallel C}}
\acro{SPMD}{\emph{Single Program Multiple Data}}
\acro{MTTI}{Mean Time Between Interrupts}
\acro{CAF}{Co-Array Fortran}
\acro{SHMEM}{Symmetric Hierarchical Memory}
\acro{FLOPS}{FLoating-point Operations Per Second} 
\acro{MPMD}{Multiple Program Multiple Data} 
\acro{SM}{Shared-Memory} 
\acro{OS}{Operating System} 
\acro{UH}{University of Houston} 
\acro{SNL}{Sandia National Laboratory} 
\acro{ANL}{Argonne National Laboratory} 
\acro{NOP}{No Operation} 
\acro{NUMA}{Non Uniform Memory Access}
\acro{BTL}{Byte Transfer Layer}
\acro{PML}{Point-to-point Management Layer}
\acro{BML}{BTL Management Layer}
\acro{MCA}{Modular Component Architecture}
\acro{DAG}{\emph{Directed Acyclic Graph}}
\acro{BCOL}{\emph{Basic Collective}}
\acro{ML}{\emph{Messaging Layer}}
\acro{P2P}{\emph{Point-to-Point}}
\acro{SM}{\emph{Shared-Memory}}
\acro{MB}{\emph{Megabyte}}
\acro{ORTE}{Open Run-time Environment}
\acro{OPAL}{Open Portable Access Layer}
\acro{IBM}{International Business Machines Corportation}
\acro{HP}{Hewlett-Packard}
\acro{SGI}{Silicon Graphics, Inc.}
\acro{DMAPP}{Distributed Memory Applications}
\acro{VM}{Virtual Machine}
\acro{TCO}{Total Cost of Ownership}
\acro{EC2}{Elastic Compute Cloud}
\acro{NERSC}{National Energy Research Scientific Computing Center}
\acro{ALCF}{Argonne Leadership Computing Facility}
\acro{MR}{\emph{MapReduce}}
\acro{NIST}{National Institute of Standards and Technology}
\acro{10GbE}{10-gigabit Ethernet}
\acro{CUDA}{\emph{Compute Unified Device Architecture}}
\acro{SMs}{Streaming Multiprocessors}
\acro{FFT}{Fast Fourier transform}
\acro{UCX}{Unified Communication X}
\acro{UCS}{UC-Services}
\acro{UCT}{UC-Transports}
\acro{UCP}{UC-Protocols}
\acro{TL}{Transport Layer}
\acro{RMA}{Remote Memory Access}
\acro{PMI}{Process Manager Interface}
\acro{SMSG}{Short message}
\acro{BTE}{Block Transfer Engine}
\acro{SSCA}{Scalable Synthetic Compact Applications}
\acro{RTE}{Run Time Environment}
\acro{SharP}{SHARed data-structure centric Programming abstraction}
\acro{HBM}{High-bandwidth Memory}
\acro{NVRAM}{non-volatile random access memory}
\acro{LLNL}{Lawerence Livermore National Laboratory}
\acro{MD}{\emph{Memory Domains}}
\acro{DT}{\emph{Data Tiers}}
\acro{UMA}{\emph{Unified Memory Allocator}}
\acro{UMI}{\emph{Unified Memory Interface}}
\acro{KV}{\emph{Key-Value}}
\acro{C/R}{\emph{Checkpoint/Restart}}
\acro{ADIOS}{Adaptable IO System}
\acro{MDS}{Metadata Server}
\acro{TEPS}{Traversed Edges Per Second}
\acro{UCX}{Unified Communication X}
\acro{UCS}{UC-Services}
\acro{UCT}{UC-Transports}
\acro{UCP}{UC-Protocols}
\acro{TL}{Transport Layer}
\acro{TLS}{Thread Local Storage}
\acro{SSCA}{Scalable Synthetic Compact Applications}
\acro{SHARP}{Mellanox Scalable Hierarchical Aggregation and Reduction Protocol}
\acro{NCCL}{NVIDIA Collective Communication Library}
\acro{RCCL}{ROCm Communication Collectives Library}
\acro{ANL}{Argonne National Laboratory}
\acro{LANL}{Los Alamos National Laboratory}
\acro{ORNL}{Oak Ridge National Laboratory}
\acro{SNL}{Sandia National Laboratory}

\acro{HCOLL}{Hierarchical Collectives}
\acro{HCA}{Host Channel Adapter}
\acro{RMA}{Remote Memory Access}
\acro{MEMIC}{Memory Mapped to InterConnect}
\acro{BAR}{Base Address Registers}
\acro{UCC}{Unified Collective Communication}
\acro{DL}{Deep Learning}
\acro{EE}{Execution Engine}
\acro{CL}{Collective Layer}
\acro{TL}{Team Layer}
\acro{DPU}{Data Processing Unit}
\end{acronym}

\begin{document}

\pagestyle{plain}  %

\title{Zero-SM NCCL: A Systematic Guide to Designing and Deploying SM-Free GPU Communication}

\newif\ifcamera
\cameratrue

\ifcamera
\author{
\IEEEauthorblockN{
Zhenhao He\IEEEauthorrefmark{2}\IEEEauthorrefmark{3},
Siyuan Shen\IEEEauthorrefmark{1}\IEEEauthorrefmark{3},
Ke Wen\IEEEauthorrefmark{2},
Prajwal Singhania\IEEEauthorrefmark{2},
Shiqing Fan\IEEEauthorrefmark{2},\\
Youngeun Kwon\IEEEauthorrefmark{2},
Sangkug Lym\IEEEauthorrefmark{2},
Vasudevan Rengasamy\IEEEauthorrefmark{2},
Vishnuswaroop Ramesh\IEEEauthorrefmark{2},\\
Yicheng Qiang\IEEEauthorrefmark{2},
Craig Belusar\IEEEauthorrefmark{2},
Jeff Hammond\IEEEauthorrefmark{2},
Torsten Hoefler\IEEEauthorrefmark{1}}
\IEEEauthorblockA{\IEEEauthorrefmark{1}ETH Zurich, Switzerland,
\{siyuan.shen, torsten.hoefler\}@inf.ethz.ch\\
\IEEEauthorrefmark{2}NVIDIA Corporation, \{zhenhaoh, kwen, psinghania, shiqingf, youngeunk, slym,\\
vrengasamy, vramesh, yqiang, cbelusar, jehammond\}@nvidia.com}
\thanks{\IEEEauthorrefmark{3}These authors contributed equally to this work.}
}
\else
\author{}
\fi

\maketitle

\begin{abstract}%
GPU communication libraries such as NCCL use Streaming Multiprocessors
(SMs) for data movement and synchronization, slowing overlapped
computation at a cost that grows with interconnect bandwidth. We present
\textbf{Zero-SM NCCL}, which communicates within and across scale-up
domains without launching \textbf{any} SM kernel. It synchronizes
through CUDA stream memory operations, moves data asynchronously with
copy engines within a scale-up domain, and drives network RDMA from a
stream-ordered CPU proxy. It supports data-movement collectives such as
AllGather and AlltoAll, plus one-sided \texttt{ncclPutSignal} and
\texttt{ncclWaitSignal} APIs. On GB200 NVL72 and B200 NVL8 systems,
Zero-SM collectives reach up to 1.36$\times$ higher AllGather bandwidth
than NCCL's legacy SM-based kernels and cut the median slowdown of
overlapped GEMMs from 33\% to 4\%, at a higher latency per operation. In
Megatron-LM training, they accelerate FSDP by up to 3.5\%, and one-sided
puts accelerate pipeline parallelism by up to 15\%. Zero-SM
communication is not a universal win, however. Through an end-to-end
study across models and parallelism strategies, we show that it pays off
for large transfers in GPU-bound workloads and for deep pipelines with
large bubbles, but loses on small messages and host-bound workloads. We
distill these findings into guidelines for enabling zero-SM
communication in NCCL and building it into other libraries.
\end{abstract}%

\section{Introduction}
\label{sec:introduction}

Large language model (LLM) training and inference run on thousands of
GPUs connected by high-bandwidth interconnects such as NVLink and
InfiniBand (IB). Every training step exchanges parameters, gradients,
and activations among these GPUs, so collective communication sits at
the heart of every parallelism strategy. Libraries such as the NVIDIA
Collective Communications Library (NCCL)~\cite{nccl} and
NVSHMEM~\cite{nvshmem,ma2026demystifying} have become the de facto runtimes for this
GPU-to-GPU data movement, and training frameworks such as
Megatron-LM~\cite{shoeybi2019megatron} rely on them for every
collective. To saturate the interconnect, these libraries run their
data transfers and synchronization as kernels on the GPU's Streaming
Multiprocessors (SMs), its general-purpose compute units.

\begin{figure}[t]
\centering
\includegraphics[width=\columnwidth]{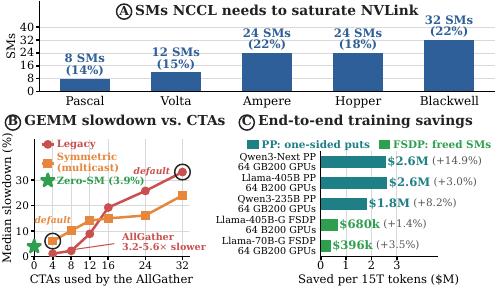}
\vspace{-2em}
\caption{\textcircled{A} SMs that NCCL needs to saturate NVLink per GPU
generation, with their share of the GPU's SMs. \textcircled{B} Median
slowdown of the Qwen3-30B-A3B GEMMs that a 144\,MiB AllGather slows by
more than 1\%, against its CTA count, on one B200 node. Rings mark default CTA counts, and the
symmetric kernel is the multicast variant. Legacy beats Zero-SM NCCL
only at 4--8 CTAs, where its AllGather is 3.2--5.6$\times$ slower.
\textcircled{C} Estimated savings of Zero-SM NCCL over a 15T-token
pretraining run, from freed SMs (FSDP) or one-sided puts (PP), with
step-time speedups in parentheses.}
\label{fig:teaser}
\vspace{-1.5em}
\end{figure}

SMs, however, are the scarcest resource in LLM training. Its general
matrix multiplications (GEMMs) launch far more thread blocks than a GPU
has SMs, so every SM held by communication delays compute. Yet the SMs
NCCL needs to saturate NVLink have grown from 8 on Pascal to 32 on
Blackwell (Figure~\ref{fig:teaser}\textcircled{A}), and SM-based kernels
can lower this cost only by giving up bandwidth
(Figure~\ref{fig:teaser}\textcircled{B}). On one B200 node, a 144\,MiB
AllGather slows the GEMMs it overlaps by about 33\% under NCCL's legacy
ring and 6\% under its symmetric multicast kernel, against 3.9\% when copy engines move the data instead of SMs. In LLM training, zero-SM communication pays
off end to end in two ways. Freeing SMs speeds up fully sharded data parallel (FSDP)
training by up to 3.5\%, and stream-ordered one-sided puts, which
decouple pipeline stages, speed up pipeline-parallel (PP) training by up
to 14.9\%. These
speedups save up to \$2.6M over a 15T-token pretraining run at on-demand
cloud prices~\cite{coreweave2026pricing}
(Figure~\ref{fig:teaser}\textcircled{C}).

Several efforts have sought to reduce the SM cost of communication. NCCL's symmetric kernels replace multi-stage pipeline protocols with single-shot algorithms~\cite{bachan2025nccl227,shen2026microsecond} that, by default, use a few CTAs instead of tens (\S\ref{sec:results:overlap}). However, SMs are still required for data movement and synchronization. NCCLX~\cite{si2026ncclx} offloads bulk data movement to dedicated hardware units but retains SM kernels for synchronization, falling short of full elimination. Moreover, most existing approaches target either scale-up or scale-out communication in isolation, leaving no SM-free option for workloads whose communication spans both.

In this work, we present \textbf{Zero-SM NCCL}, a complete solution that enables GPU communication \emph{without consuming any SMs}. Zero-SM NCCL covers both scale-up and scale-out paths and supports both collective operations (e.g., AllGather and AlltoAll) and one-sided primitives (Put, Wait) that allow users to build custom communication patterns. To achieve this, Zero-SM NCCL combines three techniques. SM-free synchronization based on CUDA stream memory operations replaces all SM-driven barriers and completion tracking. Copy engines (CEs) move data over NVLink. A CPU proxy drives network Remote Direct Memory Access (RDMA) while preserving CUDA stream semantics. Together, these techniques offload both data movement and synchronization from SMs entirely.

The contributions of this work are as follows:
\begin{itemize}[leftmargin=1em]
    \item To the best of our knowledge, Zero-SM NCCL is the first communication framework that completely eliminates SM usage for both scale-up and scale-out GPU communication, covering collective operations and one-sided primitives within a single, unified design.
    \item We introduce new one-sided APIs (\texttt{ncclPutSignal} and \texttt{ncclWaitSignal}) that enable users to build custom zero-SM communication patterns, and implement zero-SM collective algorithms including a
    hierarchical AllGather that overlaps scale-out transfers with scale-up distribution.
    \item We integrate Zero-SM NCCL into the FSDP and PP paths of Megatron-LM and evaluate it on GB200 NVL72 and B200 NVL8 systems, from microbenchmarks and a compute-overlap benchmark on GEMMs to 19 end-to-end training recipes of dense and mixture-of-experts (MoE) models.
    \item We show that zero-SM communication is not a universal win. It pays off for large transfers in GPU-bound workloads and for deep pipelines with large bubbles, but loses on small messages and host-bound workloads. We distill these conditions into guidelines that serve as a reference for enabling zero-SM communication in NCCL and for building it into other communication libraries.
\end{itemize}
\ifcamera
All code needed to reproduce our results, including the NCCL patches,
benchmark harnesses, and training recipes, is publicly
available at \artifacturl.
\fi

\section{Background}
\label{sec:background}

\subsection{GPU Interconnect Landscape}
\label{sec:bg:interconnect}

Large clusters use a two-level communication hierarchy. A
\emph{scale-up} domain connects GPUs whose memories are
load/store-accessible (LSA) to one another, through NVLink and NVSwitch or PCIe
peer access. On Blackwell, NVLink5 provides 1.8~TB/s per
GPU (900~GB/s per direction)~\cite{nvidia_nvlink}, and NVSwitch adds NVLink SHARP (NVLS), which
multicasts writes and reduces data inside the fabric. The
\emph{scale-out} network connects scale-up domains through RDMA-capable
NICs. Current DGX systems use up to eight 400~Gb/s InfiniBand or RDMA over Converged
Ethernet (RoCE) NICs per node~\cite{nvidia_dgx_b200_2025}, and per-link rates continue
to rise~\cite{connectx8, quantumx_photonics}.

\subsection{GPU Data Movement Engines}
\label{sec:bg:arch}

\noindent\textbf{GPU copy engines.}
Copy engines (CEs) serve operations such as \texttt{cudaMemcpyAsync} and
move data over NVLink without occupying SMs once a transfer is launched.
In contrast, the Tensor Memory Accelerator (TMA) offloads structured
transfers from threads~\cite{hopper2022,ptx_tma}
but still needs SM-resident kernels to issue and synchronize
them~\cite{ptx_tma}.

\noindent\textbf{RNIC DMA engines.}
RDMA-capable NICs (RNICs), such as NVIDIA ConnectX adapters, contain direct memory
access (DMA) engines that read and write GPU memory directly through GPUDirect
RDMA~\cite{nvidia2013gpudirect}. Like CEs, they move data concurrently
with SM execution, as long as the control and synchronization around each transfer
also avoid SMs.

\subsection{How NCCL Uses SMs for Communication}
\label{sec:bg:nccl}

NCCL~\cite{nccl, Zhiyi25} is the primary library for multi-GPU
communication on NVIDIA GPUs. Within a scale-up domain, its pipeline protocols (Simple, LL, and LL128)~\cite{Zhiyi25} use tens of thread
blocks, or cooperative thread arrays (CTAs), whose threads copy data
between stages, perform reductions, and poll flags. Across domains, SM
kernels pack data and post descriptors to CPU proxy threads, which drive
the RDMA operations. On both paths, SMs also perform all synchronization,
spinning on flags, executing barriers, and tracking completion. This
work is the hardest to offload, because it involves conditional control
flow that DMA engines cannot express.

\noindent\textbf{Host- vs.\ device-initiated communication.}
NCCL's collectives are \emph{host-initiated}. The CPU enqueues operations
on CUDA streams, which the GPU then executes. \emph{Device-initiated}
APIs such as NVSHMEM~\cite{nvshmem,markthub2022ibgda,ma2026demystifying} and NCCL's GPU-Initiated Networking (GIN)~\cite{nccl2.28} instead issue operations from inside CUDA kernels, which enables fusion but requires SMs. Zero-SM NCCL therefore targets the host-initiated API.

\noindent\textbf{Symmetric memory registration.}
NCCL's symmetric registration lets all ranks collectively register a
buffer, so every peer learns its address at registration
time~\cite{langer2022dynamic}. Zero-SM NCCL relies on it in three ways.
CEs copy directly into peer user buffers, NICs access them through
GPUDirect RDMA with the remote addresses that one-sided puts need, and
signal locations are registered the same way, so peers can write them
over both NVLink and the network.

\subsection{Related Work and Remaining Gap}
\label{sec:bg:prior}

Prior work reduces, but does not completely remove, the SM cost of communication.

NCCL's symmetric kernels~\cite{bachan2025nccl227,shen2026microsecond}
replace multi-stage pipelines with single-shot load/store algorithms that
need only a few CTAs, but those CTAs still move data, reduce, and poll
flags. MSCCL++~\cite{hwang2026mscclpp} exposes programmable channels whose
primitives are all invoked from GPU kernels. Even its port channels, which
hand data movement to a CPU proxy, signal and wait inside the kernel.

NCCLX~\cite{si2026ncclx} moves data with CEs and RDMA from CPU threads,
but launches a small kernel per host-initiated collective to enforce
stream order. VCCL~\cite{zhang2025vccl} offers an SM-free mode that uses
CEs and a CPU proxy with stream memory operations, but only for
point-to-point transfers. Concurrent
HOCCL~\cite{fei2026hoccl} also moves AllGather, AlltoAll, and Send/Recv
data with DMA engines, but stages cross-node data through peer GPUs,
adding an NVLink copy at each end, and is evaluated only on 32 H800
GPUs. We do not compare against HOCCL, as it appeared concurrently
with this work without public source code. On AMD GPUs, ConCCL~\cite{agrawal2025conccl},
DMA-Latte~\cite{pati2026dmalatte}, and FiCCO~\cite{pal2025ficco} run
AllGather and AlltoAll on DMA engines, and PyTorch's asynchronous tensor
parallelism~\cite{wang2024asynctp} and Flux~\cite{chang2024flux} use CE
copies for AllGather. These DMA-based designs are confined to a single node, and the fused operators still wait on SMs.

GPUDirect Async~\cite{agostini2018gpudirect} triggers pre-posted IB
operations from CUDA streams, HPE's stream-triggered Message Passing
Interface (MPI)~\cite{namashivayam2022stream} does so with Slingshot NIC triggers,
and MPIX Stream~\cite{zhou2022mpixstream} enqueues MPI operations on
streams (see~\cite{bridges2024triggering} for a survey). These target
point-to-point messages, and none provides zero-SM collectives or a
copy-engine path.

New hardware such as T3~\cite{pati2024t3},
ACE~\cite{rashidi2021ace}, and the Scalable Hierarchical Aggregation
Protocol (SHArP)~\cite{sharp:16:graham} offloads
collectives, and Maia 200~\cite{xu2026maia200} builds signaled DMA
transfers into its network protocol, sending remote semaphore updates
together with the data. Overlap systems such as
TileLink~\cite{zheng2025tilelink} and
FlashOverlap~\cite{hong2026flashoverlap} instead schedule around the
contention between compute and communication, which can slow overlapped
compute by up to 40\%~\cite{lee2025characterizing}, rather than removing it.

No existing library offers both stream-ordered collectives and one-sided
operations that launch no SM kernel at all, within and across scale-up
domains, on hardware available today. Zero-SM NCCL fills this gap
(\S\ref{sec:design}). In addition, %
we are the first to systematically analyze when zero-SM communication
pays off end to end and to distill guidelines from it.

\section{Zero-SM NCCL Design}
\label{sec:design}

\subsection{Overview and Scope}
\label{sec:design:overview}

\begin{figure}[t]
\centering
\includegraphics[width=\columnwidth]{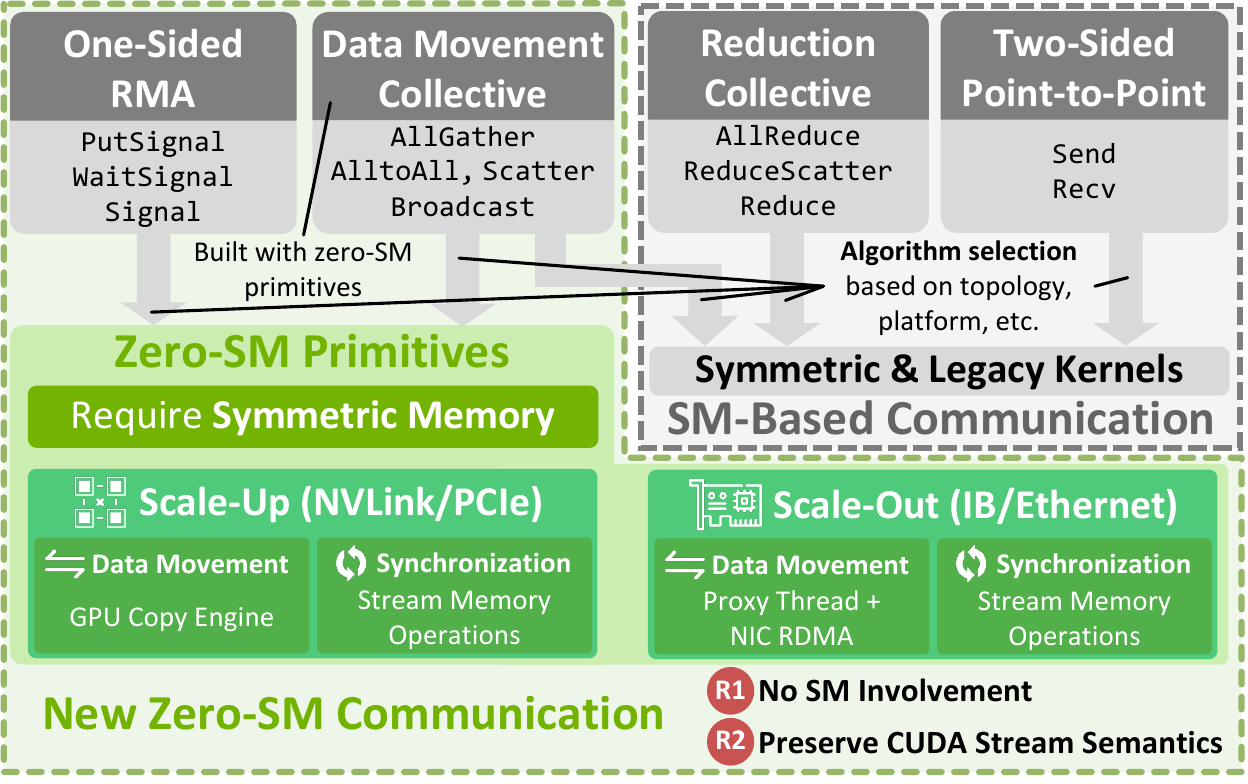}
\vspace{-2em}
\caption{Overview of Zero-SM architecture for NCCL's on-stream host API.
}
\label{fig:zero-sm-overview}
\vspace{-1.5em}
\end{figure}

Zero-SM NCCL eliminates SM consumption in GPU communication by offloading both data movement and synchronization. Figure~\ref{fig:zero-sm-overview} shows the overall architecture.

\noindent\textbf{Design requirements.}
All Zero-SM NCCL operations are designed around the following two requirements:

\noindent\circled{R1}\hspace{0.3em} \textbf{Avoid SM involvement.} A zero-SM operation must not launch GPU kernels or consume SM resources.

\noindent\circled{R2}\hspace{0.3em} \textbf{Preserve CUDA stream semantics.} Zero-SM operations must compose with other CUDA operations. They must be \emph{stream-ordered}, meaning that an operation starts only after earlier work in the same stream completes, and later work in the stream starts only after the operation completes.

\noindent\textbf{Offload strategy.}
For scale-up communication among peers, e.g., over NVLink or PCIe, data movement is offloaded to CEs. For scale-out communication over the network, data movement is driven by CPU proxy threads that post RDMA operations to the NIC. In both cases, synchronization is handled by CUDA stream memory operations (\S\ref{sec:design:sync}). All buffers and signal locations are symmetrically registered (\S\ref{sec:bg:nccl}), so senders push data directly into peer buffers with CE copies or RDMA writes, without staging copies or runtime address exchange.

\noindent\textbf{One-sided APIs.}
Zero-SM NCCL introduces two new one-sided remote memory access (RMA) APIs. \texttt{ncclPutSignal} transfers data to a remote peer and updates a remote signal upon completion, and \texttt{ncclWaitSignal} blocks the CUDA stream until a signal reaches a specified value. Both operations execute without SM kernels. These primitives allow advanced users to construct custom communication patterns (e.g., pipelined exchanges or application-specific all-to-all variants) entirely in zero-SM mode. The collectives described in \S\ref{sec:design:collectives} share the same underlying mechanisms.

\noindent\textbf{Supported collectives.}
Zero-SM NCCL targets collectives that do not involve arithmetic, namely AllGather, AlltoAll, Gather, and Scatter. These collectives are pure data movement operations and can be decomposed into pairwise transfers with synchronization, which is exactly the pattern that CEs and proxy threads can execute without SMs. Reduction collectives, such as AllReduce and ReduceScatter, require arithmetic on the data and are not yet supported in zero-SM mode.

\subsection{SM-Free Synchronization Primitives}
\label{sec:design:sync}

A sender must know that the receiver's buffer is ready, and the receiver that data has arrived. NCCL detects both with SM kernels that spin on flags (\S\ref{sec:bg:nccl}), which violates \circled{R1}. Zero-SM NCCL instead uses \emph{CUDA stream memory operations}, which are stream-ordered and do not need SMs, satisfying \circled{R1} and \circled{R2}. 
The CUDA driver's stream memory operations, e.g., \texttt{cuStreamWriteValue64} and \texttt{cuStreamWaitValue64}, write to or wait on a memory location as part of a stream without launching a kernel~\cite{stream_memory_operations}.
Zero-SM NCCL builds its synchronization from both stream memory operations and CE operations.

\noindent\textbf{Sequence-based protocol.}
To prevent race conditions, each signal location uses a monotonically increasing sequence number rather than a boolean. The sequence number starts at 0 and increments by 1 for every operation. Each wait blocks until the signal reaches the sequence number of the current operation, guaranteeing that stale values from a previous iteration cannot cause premature unblocking.

\begin{figure*}[t]
\centering
\includegraphics[width=0.95\textwidth]{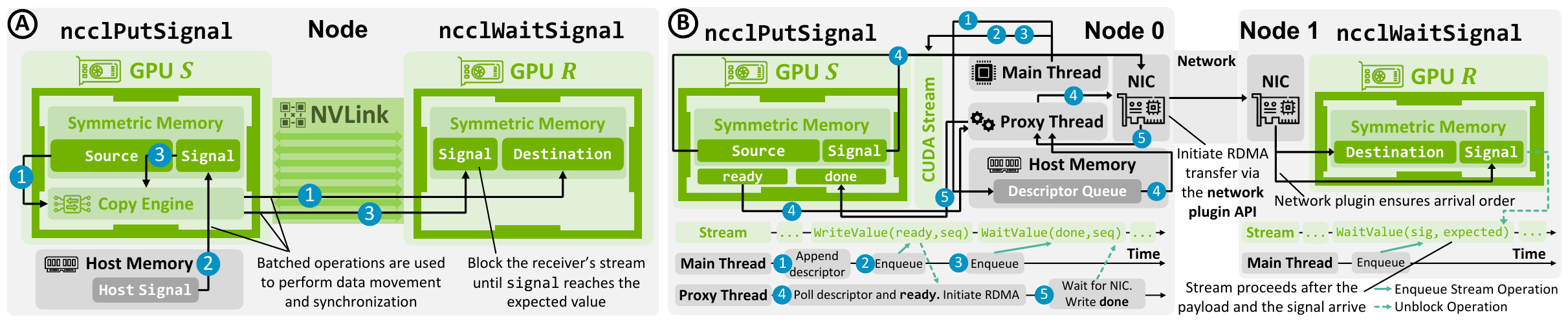}
\vspace{-1em}
\cprotect\caption{Visualization of the Zero-SM one-sided RMA protocols for \texttt{ncclPutSignal} and \texttt{ncclWaitSignal}. Panel \textcircled{A} shows the scale-up path, while Panel \textcircled{B} shows the scale-out path. Although Panel~\textcircled{A} illustrates NVLink, the same protocol also applies to PCIe and NVSwitch. The bottom of Panel \textcircled{B} provides a timeline view of the operations executed on the CUDA streams, main threads, and proxy threads of the sender and receiver.}
\label{fig:putsignal-scale-up-scale-out}
\vspace{-1.5em}
\end{figure*}

\noindent\textbf{Flag-based protocol for CUDA graph capture.}
Zero-SM NCCL also supports CUDA graph capture and replay. This requires a different synchronization protocol from ordinary stream execution. The sequence-number scheme above assumes that the host can assign a fresh expected value whenever it enqueues a new operation. During CUDA graph capture, however, the stream operations are recorded once and then replayed multiple times. The wait condition recorded in the graph cannot be changed on each replay. For instance, if the graph captured a wait for signal value $k$, and a previous replay left the signal location at value $k$ or larger, a later replay could pass the wait immediately without observing the new event.

To avoid this stale-signal problem, Zero-SM NCCL switches to replay-safe flag protocols on both the scale-up and scale-out paths when an operation is captured into a CUDA graph. Each signal location that a captured stream waits on becomes a graph-private flag that the producer sets and the consumer resets within the same replay, so every replay starts from an identical flag state and can proceed only after observing a fresh transition. This costs a few extra stream memory operations per signaled transfer for the resets and the acknowledgment. For simplicity, the remaining sections assume non-graph stream execution.

\subsection{One-Sided Operations}
\label{sec:design:onesided}

\subsubsection{Scale-Up \texttt{ncclPutSignal}}
\label{sec:design:nvlink}

Within a scale-up domain, \texttt{ncclPutSignal} moves data with the CE and synchronizes with stream memory operations. When rank~$S$ sends to rank~$R$, the main thread enqueues the following onto $S$'s stream (Figure~\ref{fig:putsignal-scale-up-scale-out}, Panel~\textcircled{A}).

\noindent\circledBlue{1}\hspace{0.3em} A device-to-device copy from $S$'s local buffer to $R$'s symmetrically registered buffer, executed by the CE.

\noindent\circledBlue{2}\hspace{0.3em} A stream memory operation that writes the current sequence number into a device-resident staging word on $S$. 

\noindent\circledBlue{3}\hspace{0.3em} A second device-to-device copy that copies the staged sequence number to $R$'s symmetrically registered signal location, notifying $R$ that the data has arrived.

Publishing the sequence number takes two operations because a stream memory operation cannot target memory on a remote GPU, leaving the CE as the only SM-free path into a peer's memory. CE copies, as currently exposed by CUDA, only move bytes and offer no atomic operation, so a copy cannot increment the destination, and the value must already sit in device memory to serve as a copy source. The host-side counter cannot be that source either, as it lives in pageable memory and copying from it would force an implicit CPU--GPU synchronization, so a stream memory operation stages the value locally as an immediate operand.

On the receiving rank~$R$, calling \texttt{ncclWaitSignal} blocks $R$'s stream until the signal reaches the expected sequence number. Because the CE transfer and the signal write are both stream-ordered, the signal is guaranteed to be written only after the data transfer completes.

\subsubsection{Scale-Out \texttt{ncclPutSignal}}
\label{sec:design:network}
Scale-out communication is more challenging because existing GPU-initiated network paths all rely on SM kernels to trigger, poll, or process network activity, which violates \circled{R1}. A purely host-driven alternative is also insufficient. Host-based network APIs, such as IB verbs or MPI one-sided operations, execute on the CPU outside CUDA stream scheduling. Used directly, they would break \circled{R2} since the CPU may start a network transfer before the GPU has finished producing the send buffer, or the GPU may consume the receive buffer before the network transfer has completed.

Zero-SM NCCL therefore coordinates the \emph{main thread} with NCCL's CPU-side proxy infrastructure, using an RMA-specific proxy progress thread to drive one-sided network operations. The main thread handles the API call, creates a descriptor for the network operation, and enqueues stream memory operations that make the descriptor visible to the proxy only after prior CUDA work completes and block later CUDA work until the transfer is done. The proxy thread then polls these descriptors, drives the actual network operation, and reports completion through the stream-visible flag.

When the sender rank invokes \texttt{ncclPutSignal}, the following sequence occurs (Figure~\ref{fig:putsignal-scale-up-scale-out}, Panel~\textcircled{B}).

\noindent\circledBlue{1}\hspace{0.3em} The main thread creates a \emph{descriptor} for the network operation, holding the source buffer, the destination window, and a monotonically increasing sequence number. It then appends the descriptor to a CPU-memory queue consumed by the proxy thread. The descriptor is associated with $\mathit{ready}$ and $\mathit{done}$ control flags, which reside in CPU-accessible memory. Depending on platform support, NCCL allocates them either as GDRCopy-mapped GPU memory or as CUDA host memory.
    
\noindent\circledBlue{2}\hspace{0.3em} The main thread enqueues \texttt{cuStreamWriteValue64} \texttt{($\mathit{ready},\mathit{seq}$)} on the stream. This write completes only after all preceding stream operations have finished, so when $\mathit{ready} \geq \mathit{seq}$, the proxy knows the source buffer contains valid data.

\noindent\circledBlue{3}\hspace{0.3em} The main thread enqueues \texttt{cuStreamWaitValue64} \texttt{(}$\mathit{done},\mathit{seq}$\texttt{)}, which blocks the stream until the proxy thread signals completion. The main thread then returns immediately.

\noindent\circledBlue{4}\hspace{0.3em} The proxy thread polls both the descriptor queue and the $\mathit{ready}$ flag. Once a descriptor is available and ready, it initiates the RDMA transfer via the network plugin API (\texttt{iput}/\texttt{iput\_signal}).

\noindent\circledBlue{5}\hspace{0.3em} The proxy thread polls for data transfer completion using the plugin's \texttt{test} function. Upon completion, it writes the operation's sequence number into the $\mathit{done}$ flag, unblocking the \texttt{cuStreamWaitValue64}.

\noindent\textbf{Signal update via network plugin backends.}
The remote signal update is exposed through NCCL's network plugin interface as an ordered put-with-signal operation (\texttt{iput\_signal}). The exact mechanism is backend-specific. An IB backend may implement the signal as an RDMA atomic (e.g., fetch-and-add) after the payload write, while other plugins can provide the same ordering semantics using their native mechanisms, including non-IB backends such as AWS Elastic Fabric Adapter (EFA). Zero-SM NCCL therefore does not depend on a particular NIC primitive. Instead, it relies on the plugin to make the signal visible after the payload transfer is complete. On the receiver, \texttt{ncclWaitSignal} enqueues a \texttt{cuStreamWaitValue64} on that signal.

\subsection{Zero-SM Collectives}
\label{sec:design:collectives}

We describe AllGather in detail, as it is the collective most commonly
used in the training workloads we target. The other zero-SM collectives are built from the same primitives and are summarized in \S\ref{sec:design:collectives:other}.
Their full implementation is available in the NCCL source
code~\cite{nccl}.

\subsubsection{Scale-Up CE AllGather}
\label{sec:design:collectives:intra}

In an $N$-GPU AllGather, each GPU contributes one chunk and receives all $N{-}1$ chunks from its peers. The protocol has three logical phases.

\noindent\textbf{Phase 1: Barrier for buffer readiness.}
Before any GPU can write into a peer's output buffer, all GPUs must have posted their buffers as ready. Zero-SM NCCL implements this barrier on symmetrically registered ready flags. Each rank publishes its ready flag to its peers with CE copies, then waits with stream memory operations until the flags from the other $N{-}1$ ranks have arrived. This ensures that every destination buffer is safe to write before the CE data transfers begin.

\noindent\textbf{Phase 2: Data broadcast.}
Once the readiness barrier is satisfied, each GPU~$i$ enqueues CE transfers to copy its local chunk into the corresponding slot of every peer's output buffer.

\noindent\textbf{Phase 3: Barrier for global completion.}
After issuing the CE transfers, Zero-SM NCCL performs a second stream-memory barrier on completion flags. This ensures that every rank has finished its transfers before any CUDA stream proceeds to subsequent work that may consume the AllGather result.

\noindent\textbf{Optimizations.}
A naive implementation of the CE path would issue many host-side CUDA calls, which can be a significant overhead for small and medium messages. Zero-SM NCCL therefore applies the following techniques.

\begin{itemize}[leftmargin=1em]
    \item \textbf{Batching CE and stream-memory operations.}
    Zero-SM NCCL batches both data movement and synchronization to reduce host-side overhead. For data movement, a collective may issue many pairwise copies, such as the $N{-}1$ remote copies. When supported, Zero-SM NCCL submits these copies through a single \texttt{cudaMemcpyBatchAsync} call. Similarly, the ready and completion barriers are grouped into batched stream-memory operations using \texttt{cuStreamBatchMemOp}. For large batches, Zero-SM NCCL inserts intermediate intra-batch synchronization after a configurable number of CE transfers to limit the amount of unsynchronized in-flight data.

    \item \textbf{Multicast.}
    When NVLS multicast is available within the local scale-up domain, Zero-SM NCCL uses it for both synchronization and data movement, as NCCL's low-latency symmetric kernels do~\cite{shen2026microsecond}. A single CE copy to a multicast address is reliably delivered by the NVLink fabric to every GPU in the local team, replacing $N{-}1$ unicast copies. When it carries data, the sender's own chunk also needs no separate self-copy. Multicast delivers to the sender as well, which lowers the effective peak bandwidth, so the tuner keeps data on multicast only below a size threshold and falls back to unicast copies for larger transfers, which hide tail latency better. Synchronization always uses multicast when available, as a barrier publishes the same flag to every peer.

    \item\textbf{Ensuring the zero-SM guarantee for intra-GPU copies.}
    Out-of-place collectives also copy each rank's own chunk within its GPU. For such same-device copies, \texttt{cudaMemcpyAsync} defaults to SMs rather than the CE, since SMs saturate HBM better. Zero-SM NCCL uses \texttt{cudaMemcpyFlagPreferOverlapWithCompute} to force them onto the CE, trading local-copy speed for the zero-SM guarantee, a loss that is often not observable because each rank's own chunk is only $1/N$ of the output and its copy runs concurrently with the peer copies.
\end{itemize}

\begin{figure*}[t]
\centering
\includegraphics[width=0.95\textwidth]{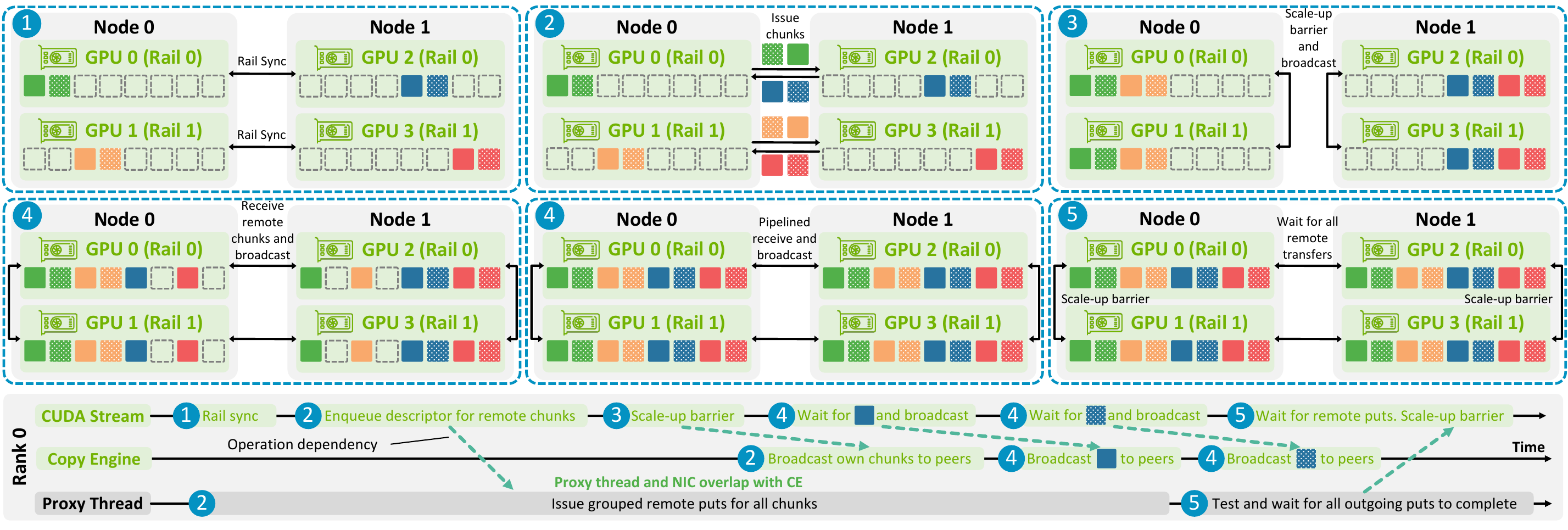}
\vspace{-1em}
\cprotect\caption{Hierarchical Zero-SM AllGather with two nodes and two GPUs per node, where each rank contributes two chunks. Solid blocks denote chunks already present in a GPU's receive buffer, while dashed blocks denote empty buffer slots. The top half shows data movement across scale-up and scale-out domains, and the bottom half shows the timeline for rank~0's CUDA stream, CE, and proxy thread. \circledBlue{1}--\circledBlue{5} link the protocol steps to \S\ref{sec:design:collectives:hier}.}
\label{fig:hierarchical-allgather}
\vspace{-1.5em}
\end{figure*}

\subsubsection{Hierarchical AllGather}
\label{sec:design:collectives:hier}

For multi-domain topologies, Zero-SM NCCL uses a hierarchical algorithm that composes the proxy-based scale-out path with the CE-based scale-up path. The implementation treats GPUs with the same local rank across scale-up domains as a \emph{rail}. Each rail exchanges data across scale-out connections using the RMA proxy, while CEs distribute the received data among GPUs within each scale-up domain. The pipeline consists of the following ordered stages, with the receive path operating at chunk granularity so that each remote chunk can be scattered locally as soon as its arrival signal is observed. Figure~\ref{fig:hierarchical-allgather} illustrates the protocol on a two-node example.

\noindent\circledBlue{1}\hspace{0.3em} \textbf{Cross-domain rail synchronization.}
Before starting data movement, each rail performs an entry barrier with signal-only RMA operations. The proxy builds a group of zero-byte put-with-signal operations to the matching local-rank peers in the other scale-up domains and waits until the corresponding inbound signals have arrived.

\noindent\circledBlue{2}\hspace{0.3em} \textbf{Grouped scale-out puts.}
Each rank splits its per-rank slice into chunks and builds a single RMA \emph{PutGroup} descriptor containing all scale-out put-with-signal operations for the remote domains. A stream-memory operation submits the group to the proxy.

\noindent\circledBlue{3}\hspace{0.3em} \textbf{Local readiness barrier and self-broadcast.}
While the proxy progresses the network puts, the CUDA stream performs a scale-up CE barrier to ensure local receive buffers are ready. Each rank then uses batched CE copies to broadcast its own local chunk into the corresponding receive-buffer slot on the other GPUs in the same scale-up domain.

\noindent\circledBlue{4}\hspace{0.3em} \textbf{Pipelined receive-and-scatter.}
For each remote domain and each chunk, the stream waits for the proxy signal indicating that the chunk has arrived from the corresponding rail peer. As soon as a chunk is available, NCCL issues batched CE copies to scatter that chunk from the rail GPU's receive buffer to the other local GPUs. This overlaps scale-out reception with scale-up distribution instead of waiting for the entire network exchange to finish. Chunks have a fixed maximum size, which trades overlap between network arrival and local scatter against per-chunk synchronization cost.

\noindent\circledBlue{5}\hspace{0.3em} \textbf{Outgoing completion and final barrier.}
After all receive-side chunk scatters have been issued, the stream waits on the PutGroup completion operations to ensure all outgoing network puts have completed. A final scale-up CE barrier then prevents later CUDA work from consuming the AllGather result until all local distribution has finished.

\noindent\textbf{Rail schedules.}
The steps above send each chunk directly to every remote domain on the
rail. For large messages and many nodes, pipelined and ring rail
schedules perform better, and a tuner selects among the three by node
count and per-rank message size (\S\ref{sec:results:micro:ag}). The
pipelined schedule and the tuner are not yet in a released NCCL version and will
ship in a future release. We do not detail these schedules, as they
follow the same zero-SM principles.

\subsubsection{Other Collectives}
\label{sec:design:collectives:other}

The other collectives reuse the AllGather structure (a readiness barrier, batched CE copies, and a completion barrier) and differ mainly in which chunks are copied where.

\section{Performance Evaluation}
\label{sec:results}

\newtcolorbox{condbox}[1]{enhanced, breakable, colback=green!3!white, colframe=green!40!black,
  boxrule=0.5pt, arc=1.5pt, left=3pt, right=3pt, top=2pt, bottom=2pt}  %
\DeclareRobustCommand{\condmark}[1]{\tikz[baseline=(c.base)]\node[circle, fill=green!40!black, text=white,
  inner sep=0.7pt, minimum size=1.15em, font=\scriptsize\bfseries](c){#1};}
\newlist{conditions}{enumerate}{1}
\setlist[conditions]{label=\condmark{\arabic*}, ref=C\arabic*, leftmargin=1.7em, labelsep=0.4em,
  itemsep=2pt, topsep=2pt, parsep=0pt}

We perform a comprehensive evaluation of Zero-SM NCCL in three steps. Microbenchmarks measure raw communication performance against NCCL's SM-based kernels and against alternative libraries (\S\ref{sec:results:micro}). An overlap benchmark measures how much each implementation slows the compute it runs next to (\S\ref{sec:results:overlap}). Finally, end-to-end Megatron-LM training on FSDP and PP recipes shows when these effects turn into faster training, and when they do not (\S\ref{sec:results:e2e}).

\subsection{Experimental Setup}
\label{sec:results:setup}

\begin{table}[t]
\centering
\footnotesize
\caption{Platforms used in the evaluation.}
\vspace{-1em}
\label{tab:eval-setup}
\begin{tabular}{@{}lll@{}}
\toprule
 & \textbf{GB200 NVL72} & \textbf{B200 NVL8} \\
\midrule
GPUs per node (SMs)  & GB200: 4 (148)             & B200: 8 (148) \\
NVLink domain        & 18 nodes (72 GPUs)      & 1 node (8 GPUs) \\
NVLink               & NVLink5, 900\,GB/s/dir  & NVLink5, 900\,GB/s/dir \\
Scale-out NICs       & 4$\times$ 400\,Gb/s per node & 8$\times$ 400\,Gb/s per node \\
CPU                  & Grace (aarch64)         & 2$\times$ Xeon Platinum 8570 \\
Driver               & 580.178.04              & 580.167.08 \\
\bottomrule
\end{tabular}
\vspace{-2em}
\end{table}

\begin{figure}[t]
\centering
\includegraphics[width=\columnwidth]{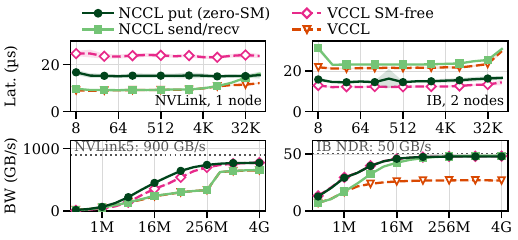}
\vspace{-2em}
\caption{Point-to-point latency (top) and bandwidth (bottom) by message
size on B200 NVL8, within a node over NVLink (left) and across two nodes
over one 400\,Gb/s rail (right). At bulk sizes, the zero-SM put matches
or exceeds the SM-based Send/Recv. Bands show $\pm$1 standard deviation
over 10 trials.}
\label{fig:micro-p2p}
\vspace{-1.5em}
\end{figure}

\begin{figure*}[t]
\centering
\includegraphics[width=\textwidth]{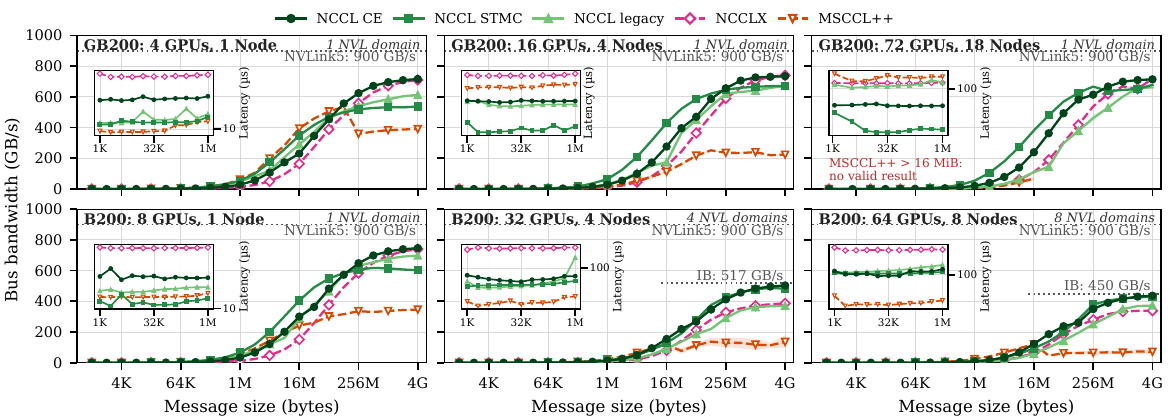}
\vspace{-2em}
\caption{AllGather bus bandwidth and latency. Colored bands show $\pm$1 standard deviation over 10 trials.}
\label{fig:micro-ag}
\vspace{-1.5em}
\end{figure*}

Table~\ref{tab:eval-setup} lists the two systems. They sit at opposite
ends of the scale-up design space. On the GB200 NVL72 system, an entire
18-node rack forms one NVLink domain, so every experiment of up to
72~GPUs stays on the CE path. On the B200 NVL8 system,
each node is its own NVLink domain, so every multi-node collective takes
the hierarchical path of \S\ref{sec:design:collectives:hier}. All
experiments are based on NCCL~2.32.3, extended with some features that
are not yet released, such as the pipelined rail schedule
(\S\ref{sec:design:collectives:hier}).
\ifcamera
The patches, run scripts, and recipes for every experiment are in our
artifact repository~\artifacturl.
\fi

\subsection{Microbenchmarks}
\label{sec:results:micro}

\noindent\textbf{Setup.}
We use NCCL's performance tests, \texttt{nccl-tests}~\cite{nccl-tests}, with 10
trials per data point, and sweep message sizes from 8\,B (point-to-point) or 1\,KiB (AllGather) to 4\,GiB. Within NCCL, we compare three implementations, namely the legacy SM-based ring kernels, the symmetric-memory kernels~\cite{bachan2025nccl227,shen2026microsecond}, and Zero-SM NCCL (\texttt{NCCL\_CTA\_POLICY=ZERO}), labeled CE in the results. NCCL's default tuner selects the protocol (e.g., LL or Simple) and the symmetric algorithm (the unicast ST or multicast STMC kernel) for each implementation.

We also compare against three alternative libraries described in
\S\ref{sec:bg:prior}, each of which moves part of the communication off
the SMs, namely VCCL~\cite{zhang2025vccl} in both its default and
SM-free point-to-point modes, NCCLX's persistent AllGather
(\texttt{AllGatherP})~\cite{si2026ncclx,torchcomms}, which moves data
with CEs, and MSCCL++~\cite{hwang2026mscclpp}, whose port channels hand
inter-node transfers to a CPU proxy while SM kernels move all
intra-node data.

We tuned NCCLX and MSCCL++ for each GPU count and node layout. For NCCLX we select the best of its three \texttt{AllGatherP} algorithms, and for MSCCL++ we grid-search the thread-block count, threads, and protocol and pick the switch point between its LL and Simple protocols.

\subsubsection{Point-to-Point Bandwidth and Latency}
\label{sec:results:micro:p2p}

Figure~\ref{fig:micro-p2p} compares the zero-SM one-sided put with two-sided Send/Recv and with VCCL. Over NVLink, the CE-based put has a higher latency than the Send/Recv kernel (16.7 vs.\ 9.7\,$\mu$s at 8\,B). It overtakes the kernel from 64\,KiB, and ends at 773 vs.\ 658\,GB/s at 4\,GiB. VCCL's SM-free mode reaches a similar bulk bandwidth (774\,GB/s) with a latency floor of about 25\,$\mu$s. Across nodes, the proxy-driven put halves the latency of Send/Recv (15.9 vs.\ 31.4\,$\mu$s at 8\,B), VCCL's SM-free mode is slightly faster still (13.0\,$\mu$s), and all three saturate the 50\,GB/s rail, while default VCCL reaches only 27\,GB/s.

\subsubsection{AllGather Bandwidth and Latency}
\label{sec:results:micro:ag}

We focus on AllGather (Fig.~\ref{fig:micro-ag}), the collective most
commonly used in the training workloads we target.

\noindent\textbf{Within one NVLink domain,} CE is the fastest NCCL implementation for bulk transfers. At 4\,GiB it reaches 714--746 GB/s, against 537--678 GB/s for symmetric and 613--698 GB/s for legacy. CE outperforms the SM kernels at bulk sizes because on Blackwell a CE issues wider NVLink transactions than SM-initiated copies. The price is a
higher latency floor. Every CE collective is issued from the host and
synchronized through stream memory operations, so a 1\,KiB AllGather
takes 25--29\,$\mu$s on one node, against 12--18\,$\mu$s for the SM
kernels. For mid-sized messages the symmetric kernel therefore remains
faster, and CE overtakes it only from 128--512\,MiB. At 72~GPUs the CE
floor grows to 52\,$\mu$s, still less than half of legacy's
118\,$\mu$s, because a ring's latency grows with the number of ranks.

\noindent\textbf{Across NVLink domains} (B200, 4 and 8 nodes), the
network sets the limit. With the pipelined rail schedule, CE reaches
503 and 436\,GB/s at 4\,GiB, close to the NIC bound (517 and 450\,GB/s).
This bound exceeds a node's 400\,GB/s of NIC bandwidth because only the
$(N{-}1)/N$ of the data from other nodes crosses the network, which
bounds bus bandwidth at $400\cdot\frac{N}{N-1}\cdot\frac{n-1}{n}$\,GB/s
for $N$ nodes and $n$ GPUs. CE is thus on par with the symmetric GIN kernel (482 and 430\,GB/s), while
legacy saturates at 372\,GB/s.

\noindent\textbf{Alternative libraries.} NCCLX's CE-based AllGatherP
matches CE at bulk sizes within one NVLink domain, but takes 60--239\,$\mu$s per call and reaches only 387 and
339\,GB/s across B200 nodes. MSCCL++ instead spends SMs. Its intra-node
kernels use up to 35 CTAs of 1,024 threads, which gives it the lowest
small-message latency on 4 GB200 GPUs and across B200 nodes and the
highest bandwidth from 1 to 64\,MiB on 4 GB200 GPUs. Above 32\,MiB per
rank, its NCCL-API layer switches to a slower full-mesh algorithm (it
would normally defer to NCCL, which we disallow), and at 72~GPUs it
produced no valid result beyond 16\,MiB.

In summary, for raw performance, zero-SM communication matches or exceeds NCCL's SM-based kernels and the alternative libraries at bulk sizes without using any SMs, while SM-based designs keep the edge for small and medium messages.

\subsection{Overlap Benchmark}
\label{sec:results:overlap}

\begin{figure}[t]
\centering
\includegraphics[width=\columnwidth]{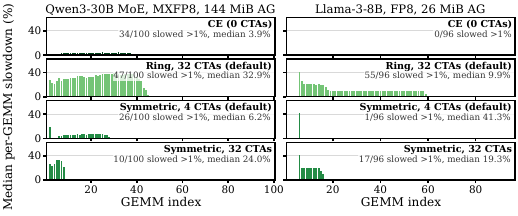}
\vspace{-2em}
\caption{Per-GEMM slowdown along a compute chain with one concurrent
AllGather. Each bar is one GEMM (median over paired
observations).}
\label{fig:overlap-timeline}
\vspace{-1.5em}
\end{figure}

\noindent\textbf{Setup.}
The overlap benchmark measures how much a concurrent collective operation slows
the compute it overlaps, one GEMM at a time. We replay GEMM chains taken from
\emph{production pretraining traces}, with each GEMM's exact shape, kernel, and
grid, on one B200 NVL8 node:
\begin{itemize}[leftmargin=1em]
  \item \emph{Qwen3-30B-A3B}: 100 MoE GEMMs in microscaling FP8 (MXFP8), overlapped with a
  144\,MiB/rank AllGather.
  \item \emph{Llama-3-8B}: 32 FP8 forward, data-gradient, and
  weight-gradient GEMM triplets (96 GEMMs), overlapped with a
  26\,MiB/rank AllGather.
\end{itemize}
Each GEMM launches several hundred thread blocks (640 and 512 for the
Qwen and Llama kernels), far more than the
148~SMs of a B200. The GEMMs thus saturate the GPU, which makes the
cost of SM-based communication as visible as possible.
We replay whole chains rather than single GEMMs for two reasons. The first version of the benchmark paired one GEMM with one AllGather, but these GEMMs run for
only tens of microseconds, far shorter than the AllGather, so launch
jitter decided whether the two overlapped at all. A chain removes this
dependence on alignment, since the AllGather always overlaps some 
of it. We also observed that a GEMM behaves differently depending on
which phase of the collective it overlaps, and a chain exposes the start,
steady state, and end of the collective in a single run.

The whole chain is enqueued first. An event recorded before a fixed GEMM
releases the AllGather on a second stream. We time every GEMM with the CUDA Profiling Tools Interface (CUPTI) in a clean pass and
in an overlap pass, pair the two by process, rank, trial, and chain
position, and report the median slowdown over four 
runs of 8~ranks $\times$ 30 trials. We compare four AllGather
configurations, namely CE, legacy ring at its default 32~CTAs, and the
symmetric kernel at its default 4~CTAs and forced to 32~CTAs.

\noindent\textbf{Results.}
Figure~\ref{fig:overlap-timeline} shows the slowdown of every GEMM in
the chain. The SM kernels slow the GEMMs they overlap in proportion to
the SMs they occupy. With the Qwen AllGather, legacy NCCL
slows the GEMMs by a median of 32.9\% and the symmetric kernel by 6.2\% at its default 4 CTAs and 24.0\% at 32 CTAs. More CTAs thus finish the AllGather sooner
but hit each overlapped GEMM harder. CE slows them by only 3.9\%, a residual that comes from contention for memory. With the smaller Llama AllGather, CE slows no GEMM by more than 0.2\%, while legacy slows them by a median of 9.9\%.

\begin{table}[t]
\centering
\scriptsize
\caption{End-to-end Megatron-LM pretraining recipes. G and H mark GPU-
and host-bound recipes. DP, TP, PP, VPP, CP, and EP are the data-,
tensor-, pipeline-, virtual-pipeline-, context-, and expert-parallel
sizes, MBS and GBS the micro- and global batch sizes, and GA the number
of gradient-accumulation steps, $GA=GBS/(MBS\cdot DP)$. FSDP recipes have
PP$=$1.}
\vspace{-1em}
\label{tab:e2e-recipes}
\setlength{\tabcolsep}{2.0pt}
\begin{tabular}{@{}cllrlrrr@{}}
\toprule
 & \textbf{Recipe} & \textbf{Type} & \textbf{GPUs} & \textbf{Parallelism} & \textbf{MBS/GBS} & \textbf{GA} & \textbf{Seq} \\
\midrule
\multirow{12}{*}{\rotatebox{90}{\textbf{FSDP}}}
 & Llama-405B-G$^{*}$ & Dense  & 64 & DP32 TP2      & 1/64  & 2 & 8K \\
 & Llama-70B-G        & Dense  & 64 & DP64 TP1      & 1/128 & 2 & 8K \\
 & Llama-70B-H$^{\ddagger}$  & Dense  & 64 & DP32 TP2      & 1/64  & 2 & 8K \\
 & Qwen3-32B-G1       & Dense  & 64 & DP32 TP2      & 8/256 & 1 & 4K \\
 & Qwen3-32B-G2       & Dense  & 32 & DP16 TP2      & 4/64  & 1 & 4K \\
 & Qwen3-32B-H        & Dense  & 64 & DP32 TP2      & 1/64  & 2 & 4K \\
 & Nemotron-15B-G1       & Dense  & 32 & DP32 TP1      & 4/256 & 2 & 4K \\
 & Nemotron-15B-G2       & Dense  & 32 & DP8 TP4       & 8/128 & 2 & 4K \\
 & Nemotron-15B-H        & Dense  & 32 & DP8 TP4       & 2/32  & 2 & 4K \\
 & Qwen3-235B-G       & MoE    & 64 & DP64 TP1 EP2  & 1/64  & 1 & 16K \\
 & Qwen3-Next-H       & Hybrid & 64 & DP64 TP1 EP2  & 1/64  & 1 & 4K \\
 & DeepSeek-V3-G$^{\dagger}$ & MoE    & 32 & DP32 TP1 EP4  & 1/64  & 2 & 4K \\
\midrule
\multirow{7}{*}{\rotatebox{90}{\textbf{PP}}}
 & Llama-405B           & Dense  & 64 & DP2 TP4 PP8 VPP2      & 1/64 & 32 & 8K \\
 & Llama-70B            & Dense  & 64 & DP2 TP4 PP4 VPP5 CP2  & 1/64 & 32 & 8K \\
 & Qwen3-32B            & Dense  & 64 & DP4 TP8 PP2 VPP2      & 2/64 & 8 & 4K \\
 & Nemotron-15B            & Dense  & 32 & DP4 TP4 PP2 VPP2      & 2/32 & 4 & 4K \\
 & Qwen3-235B           & MoE    & 64 & DP8 TP1 PP8 VPP2 EP8  & 1/64 & 8 & 4K \\
 & Qwen3-Next           & Hybrid & 64 & DP8 TP1 PP8 VPP2 EP8  & 1/64 & 8 & 4K \\
 & DeepSeek-V3$^{\dagger}$     & MoE    & 32 & DP4 TP1 PP8 VPP4 EP4  & 1/64 & 16 & 4K \\
\bottomrule
\end{tabular}\\[2pt]
\parbox{\columnwidth}{\scriptsize $^{*}$96-layer shape on B200 (126 layers on GB200).
$^{\dagger}$16-expert proxy with MXFP8 experts and full-iteration CUDA graphs.
$^{\ddagger}$GPU-bound on B200. Qwen3-Next is a hybrid of linear and full attention with MoE.}
\vspace{-2em}
\end{table}

The timelines also show the inherent limitations of such benchmarks. The
slowdown of a GEMM depends not only on how many SMs the collective holds
but on \emph{which phase} of the collective it overlaps. The GEMM in
which the symmetric AllGather starts is hit hardest, the
steady-state phase costs far less, and different GEMM types in the same
chain react differently. In real training, the
alignment between collective phases and compute kernels shifts from one
iteration to the next, and several collectives and kernel types overlap at
once. A per-GEMM slowdown measured in isolation therefore cannot predict the
end-to-end effect, which we measure next.

\subsection{End-to-End Training}
\label{sec:results:e2e}

\noindent\textbf{Integration into Megatron-LM.}
We integrate Zero-SM NCCL into Megatron-LM~\cite{shoeybi2019megatron}
starting with two communication patterns that exercise both of its
collective and one-sided put interfaces, namely the parameter AllGathers of FSDP and the stage-to-stage transfers of PP. Other
collectives, such as the AlltoAll of expert parallelism (EP), can likewise be
replaced by their zero-SM counterparts. We focus on FSDP and PP as both already have upstream adoption in Megatron-LM.

For \emph{FSDP}, Megatron-FSDP already allocates its parameter
and gradient buffers from an \texttt{ncclMemAlloc}-backed pool when NCCL
user buffers are enabled. We register this pool as a symmetric window,
so the parameter AllGathers become eligible for the zero-SM path without any
change to the FSDP schedule. Because FSDP issues AllGathers back to back,
we extend NCCL to pack consecutive zero-SM AllGathers into one plan that
shares a single synchronization envelope, amortizing the host issue and
completion cost per collective.

For \emph{PP}, we replace the stage-to-stage \texttt{ncclSend} and \texttt{ncclRecv} with PyTorch symmetric memory's
\texttt{put\_signal} and \texttt{wait\_signal}, which map to
\texttt{ncclPutSignal} and \texttt{ncclWaitSignal}. Each stage writes into a
ring of receive slots on its peer, sized by the pipeline warm-up depth,
and the receiver consumes the symmetric buffer in place.

\noindent\textbf{Recipes.} Table~\ref{tab:e2e-recipes} lists the 12
FSDP and 7 PP recipes, covering dense, MoE, and hybrid
models from 15B to 405B parameters on 32 or 64~GPUs. They are a small
subset of the possible configurations, chosen to span model types
and parallel layouts, which is enough to show
when zero-SM NCCL would be beneficial. We run FSDP and
PP separately rather than combined, to isolate the
effect of the zero-SM AllGather from that of one-sided puts. We
classify each recipe as \emph{GPU-bound} or \emph{host-bound},
depending on whether the host runs far enough ahead that the GPU never waits for work to be launched.

\noindent\textbf{Arms.} In the FSDP recipes, only the AllGather
implementation changes. It is the legacy ring, symmetric with
unicast (ST) or multicast (STMC), or CE with unicast or
multicast.
On B200 we report only the multicast arms, because the choice between unicast and multicast had a negligible effect on step time on all recipes. The two TP4
Nemotron-15B recipes have no symmetric arm on B200. In this layout, with
two data-parallel ranks per node, every AllGather of the symmetric arm
still ran the legacy ring kernel, so we omit it. In the PP recipes, one-sided puts are compared with two-sided Send/Recv.

We did not run the alternative libraries of \S\ref{sec:results:micro} as their low-SM paths are not drop-in replacements for Megatron-LM's \texttt{ncclAllGather} calls. Integrating them would change more than the AllGather implementation, so a difference in step time could no longer be attributed to the AllGather alone.

\noindent\textbf{Protocol.} Differences between node allocations on these
systems (1--5\%) are non-negligible, so we never
compare runs across allocations. Instead, arms are compared in
\emph{paired blocks}. A block runs one session per arm back to back on the same
nodes and in rotated order, with warm-up iterations discarded. Each
block yields a speedup $T_\text{baseline}/T_\text{arm}-1$ from mean step
times $T$, and we report the mean and standard deviation over three
blocks per recipe.

\begin{figure}[t]
\centering
\includegraphics[width=\columnwidth]{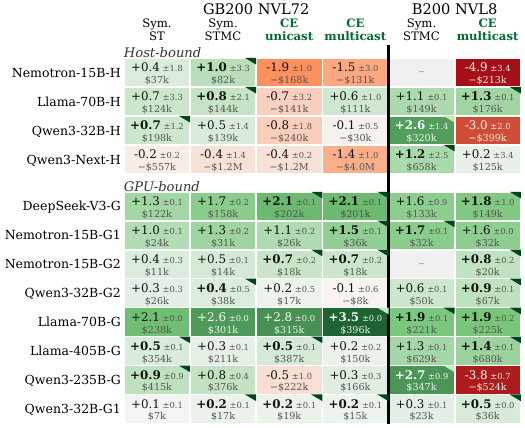}
\vspace{-2em}
\caption{FSDP step-time speedup over the legacy ring AllGather (\%, mean
and $\pm$1 standard deviation over paired blocks). Rows are grouped by
their bottleneck. A corner mark and a bold value flag the best
arm. The second line of each cell is the estimated saving
over a 15T-token pretraining run.}
\label{fig:e2e-fsdp}
\vspace{-1.5em}
\end{figure}

\noindent\textbf{A host-side artifact.} While profiling, we found a
platform effect on GB200 that penalizes host-bound training in every
arm. Host-side memory unmapping waits on GPU address-translation
invalidations, and far longer when buffers are mapped to peer GPUs and
NVLink traffic is in flight. We remove it with a small preloaded
allocator shim that routes the Python interpreter's memory arenas through
the C library's \texttt{malloc} and \texttt{free}. Freed arenas are then
reused instead of unmapped, which removes these unmaps from the training
loop. The shim is active in every arm on both systems, so it does not bias any comparison.
It is a temporary workaround, and a more robust, permanent fix is under development.

\noindent\textbf{Cost estimate.} Step-time gains of a few percent
translate into large savings over a full training run. We convert each speedup into dollars with CoreWeave's on-demand
pricing~\cite{coreweave2026pricing}, \$10.50 per GB200 GPU-hour and
\$8.60 per B200 GPU-hour. For a recipe on $N$ GPUs with price $p$ per
GPU-hour, legacy step time $T$, and $G$ tokens per step (GBS times
sequence length), training costs $pNT/G$ per token. An arm with speedup
$s$ saves the fraction $1-1/(1+s)$ of that cost, which we scale to a
pretraining run of 15T tokens, the budget of Llama~3~\cite{grattafiori2024llama3}.
Note that this is not a full deployment cost model. Production runs use more
GPUs and different layouts, and their costs depend on factors we do not
model.

\subsubsection{FSDP Results}
\label{sec:results:e2e:fsdp}

Figure~\ref{fig:e2e-fsdp} shows the FSDP results. Against the legacy
ring, CE speeds up the GPU-bound dense recipes on both systems, by up
to 3.5\% on GB200 and 1.9\% on B200 (the one exception, Qwen3-32B-G2, is within noise), because the ring kernel holds its
SMs for the entire duration of every parameter AllGather. Against the
symmetric kernel, which uses only a few CTAs, the picture is mixed. CE
wins clearly only on Llama-70B-G on GB200 and narrowly on
Nemotron-15B-G1 and the MoE DeepSeek-V3-G, ties or wins narrowly on the
other GPU-bound dense recipes, and loses on host-bound recipes and on
Qwen3-235B-G. On Qwen3-235B-G (B200), CE loses because plan fusion does not
yet cover the faster rail schedules of \S\ref{sec:design:collectives:hier}. Combining these results with profiles of each recipe, we identify three
conditions under which zero-SM collectives pay off, each of which
matters only once the previous ones hold.

\begin{condbox}{When do zero-SM collectives pay off?}
\begin{conditions}
  \item \label{cond:fsdp:cost}\textbf{The step is GPU-bound, so the freed
  SMs outweigh the extra host work.} Communication kernels slow the compute they overlap by
  occupying SMs, by 3--4\% in our end-to-end profiles.
  CE removes this cost but adds host-side work per collective. Issuing one
  CE AllGather takes 26--41\,$\mu$s on GB200,
  about 2--3$\times$ the 11--17\,$\mu$s of launching the symmetric kernel,
  and 77--86\,$\mu$s against 5--7\,$\mu$s on B200. This work stays off the critical path only
  when the step is GPU-bound. 
  
  \item \label{cond:fsdp:size}\textbf{Messages are large enough.} CE has a
  higher fixed cost per collective than SM kernels, a latency floor of
  25--29\,$\mu$s against 12--18\,$\mu$s on one node, and it overtakes the
  symmetric kernel in bandwidth only from 128--512\,MiB
  (\S\ref{sec:results:micro:ag}). Training shows the same threshold. On
  GB200, FSDP gathers of up to about 240\,MiB tie or lose against the
  symmetric kernel, and 820\,MiB wins.
  \item \label{cond:fsdp:traffic}\textbf{The route matches the
  concurrent traffic.} The choice between multicast and unicast depends on
  how the competing traffic is spread, for CE and the symmetric kernels
  alike. A multicast copy is a single stream that injects one copy but advances at the pace of its most congested destination.
  It therefore wins when links are idle or more evenly loaded, as under the
  expert all-to-all of MoE recipes. It loses when
  the load concentrates on a few peers, as under concurrent TP traffic on GB200, where congestion on one link stalls the whole AllGather while
  independent unicast copies proceed.
\end{conditions}
\end{condbox}

These conditions also show that the SMs a collective frees are only part
of the story. The kernels interfere with other communication as much as
with compute. Under concurrent all-to-all, a unicast CE AllGather on
GB200 is delayed by 1.8\,ms, six times more than its multicast
counterpart, while on B200 all-to-all delays the legacy ring by up to
8.2\,ms per AllGather and CE by at most 1.3\,ms. Choosing a
communication path therefore means accounting for its effect on both the
compute and the other collectives it overlaps.

\subsubsection{PP Results}
\label{sec:results:e2e:pp}

\begin{figure}[t]
\centering
\includegraphics[width=\columnwidth]{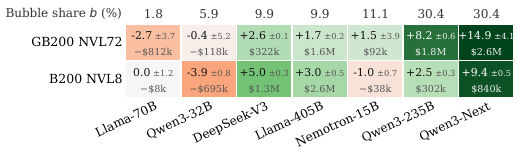}
\vspace{-2em}
\caption{PP step-time speedup of one-sided puts over
two-sided Send/Recv (\%, mean and $\pm$1 standard deviation over paired
blocks), for the PP recipes of Table~\ref{tab:e2e-recipes} ordered by
pipeline-bubble share $b$. The second line of each cell is the estimated
saving over a 15T-token training run.}
\label{fig:e2e-pp}
\vspace{-1.5em}
\end{figure}

Figure~\ref{fig:e2e-pp} shows that one-sided puts speed up every deep
pipeline on both systems, by up to 14.9\% (Qwen3-Next on GB200). Shallow or heavily interleaved
pipelines gain nothing or lose, namely Llama-70B ($b\approx 2\%$) and
the two-stage Qwen3-32B and Nemotron-15B.

The freed SMs play a minor role, speeding up compute by at most
1--2\%. In this case, the gains come primarily from removing the
rendezvous of Send/Recv.
Pipeline stages rarely finish a micro-batch together, because of uneven
layer splits, the embedding and loss on the end stages, and varying
expert loads in MoE models. With Send/Recv, a fast stage waits for its
slower neighbor at every boundary. A one-sided put writes into a
preallocated slot and moves on, so only the receiver waits, and only
when it needs the data. The gain therefore grows with the time spent in
the pipeline bubble, measured by the bubble share
$b=(PP-1)/(GA\cdot VPP+PP-1)$.

These savings come at a cost. Every put is issued from the host and
synchronized with stream memory operations, adding a fixed cost per
transfer that a Send/Recv kernel avoids. The sender must also copy each
activation or gradient into a registered staging buffer, since
Megatron-LM allocates them in ordinary PyTorch memory. Whether one-sided
puts pay off thus depends on how much waiting they remove relative to
these per-transfer costs, which leads to two conditions.

\begin{condbox}{When do one-sided puts pay off?}
\begin{conditions}
  \item \label{cond:pp:bubble}\textbf{There is waiting to remove.} The
  pipeline must be deep, with stages that drift apart, and spend much of
  the step in its bubble ($PP\ge 4$ and $b\gtrsim 10\%$). At $b=30\%$ the
  speedup is 8--15\% on GB200. A two-stage pipeline has a single boundary
  and little stage-to-stage variance, so the one-sided cost is paid in
  full, and Qwen3-32B loses 3.9\% on B200.
  \item \label{cond:pp:calls}\textbf{Transfers are few and large}, so
  that the fixed cost per put and the staging copy stay small compared
  with the waiting time removed. With deep interleaving (Llama-70B,
  $b\approx 2\%$, 322 transfers per iteration), this cost cancels the
  gain.
\end{conditions}
\end{condbox}

\subsubsection{Summary}
\label{sec:results:e2e:summary}

\textbf{When a workload meets the applicable conditions above, zero-SM
NCCL delivers substantial gains. It is
not, however, a silver bullet, and enabling it does not by itself make
training faster.} Real workloads are far more complex than
microbenchmarks suggest. Whether zero-SM communication pays off depends
not only on its raw performance, but also on
how it interacts with everything it overlaps, from compute kernels to
other collectives, and on how much host work it adds to the step. The
resulting step-time differences range from a few percent on FSDP to
nearly 15\% on deep pipelines, and over a full
pretraining run they amount to hundreds of thousands to millions of
dollars. Mapping these effects required careful end-to-end measurement,
about 2,800 GPU-hours for the reported FSDP and PP runs alone, across two contrasting systems and 19 recipes.
We hope the conditions distilled from these measurements provide a reference
for deciding when to enable zero-SM communication in NCCL, and for
designing it into other libraries.

\section{Discussion}
\label{sec:discussion}

\noindent\textbf{Other workloads.}
Weight synchronization in asynchronous reinforcement learning
(RL)~\cite{sheng2025hybridflow,wu2025llamarl} is a prime fit, a bulk
transfer that overlaps SM-bound generation.
NCCL-M2N~\cite{nccl-m2n}, which reshards weights from trainers to
inference engines with one-sided transfers, could benefit greatly from
Zero-SM NCCL in frameworks such as veRL~\cite{sheng2025hybridflow},
leaving every SM to generation during an update. Expert-weight
prefetching~\cite{li2026dwdp}, KV-cache transfer in disaggregated
serving~\cite{patel2024splitwise,zhong2024distserve}, and ring
attention~\cite{liu2024ringattention} share this pattern, and
attention-FFN disaggregation already drives its M-to-N transfers from
the CPU to keep SMs free~\cite{zhu2025megascaleinfer,stepfun2025step3}.

\noindent\textbf{Applicability beyond NVIDIA GPUs.}
The design needs only stream-ordered memory waits and writes, SM-free
copy engines, and NIC access to registered GPU memory. AMD HIP, for
instance, provides stream memory operations, peer access, and batched
asynchronous
copies~\cite{rocm_hip_stream_memory,rocm_hip_peer_access,rocm_hip_memory}.

\noindent\textbf{Reductions via NVLink SHARP.}
Reductions such as AllReduce need arithmetic, but NVLS
(\S\ref{sec:bg:interconnect}) performs it inside the NVLink fabric and
could make them zero-SM.

\noindent\textbf{Limitations.}
Within a scale-up domain, Zero-SM NCCL has higher latency than SM
kernels for small messages, since every operation is issued from the
host and synchronized through stream memory operations
(\S\ref{sec:results:micro:ag}). Graph capture costs even more, because
its replay-safe protocol needs extra signaling (\S\ref{sec:design:sync}).
Better CUDA support, such as CE-based atomics for signaling, would
reduce both costs. For scale-out synchronization, the proxy path
relies on a CPU thread, which consumes host resources. The number of CEs on the GPU may also limit
concurrent transfers with high fan-out.

\section{Conclusion}
\label{sec:conclusion}

Communication libraries pay for bandwidth with the GPU's compute,
taking more SMs with every faster interconnect. Zero-SM NCCL breaks this
link, communicating within and across scale-up domains without a single
SM kernel. In Megatron-LM, it speeds up FSDP by up to 3.5\% by freeing
SMs and PP by up to 14.9\% through one-sided puts. These gains are not
automatic. We spent about 2,800 GPU-hours of end-to-end training to
learn when they hold, and we share the resulting guidelines so that
others can adopt zero-SM communication without repeating this effort.

\ifcamera
\section*{Acknowledgments}

The authors thank Maximilian Heer for insightful discussions.
The research was conducted as part of the FastTrackAI project at the Singapore-ETH Centre, which was established
collaboratively between ETH Zurich and the National Research Foundation, Singapore. This research is supported by
the National Research Foundation, Singapore (NRF), and the Ministry of Digital Development and Information (MDDI)
under the AI Visiting Professorship (Award No.~\mbox{AIVP-2025-005}). This work also received funding from the European
Research Council (Project PSAP, No.~101002047).  The authors used GPT-6 Sol~\cite{openai2026gpt6sol} and Claude Opus 5.5~\cite{anthropic2026opus55} to assist with light editing and proofreading. All content and ideas remain the original work of the authors.

\fi

\bibliographystyle{IEEEtran}

\bibliography{references/zero_sm}

\end{document}